\documentclass{aa}  

\usepackage{graphicx}
\usepackage{txfonts}
\usepackage{hyperref}
\begin{document} 

\title{Ups and downs in the X-ray emission of the colliding wind binaries HD\,168112 and HD\,167971\thanks{Based on data collected with {\it XMM-Newton}, an ESA Science Mission with instruments and contributions directly funded by the ESA Member States and the USA (NASA).}}
\author{G.\ Rauw\inst{1} \and R.\ Blomme\inst{2} \and Y.\ Naz\'e\inst{1}\thanks{F.R.S.\-FNRS Senior Research Associate} \and D.\ Volpi\inst{2,3} \and S.\ Fernandez-Vera\inst{1}}
\institute{Space sciences, Technologies and Astrophysics Research (STAR) Institute, Universit\'e de Li\`ege, All\'ee du 6 Ao\^ut, 19c, B\^at B5c, 4000 Li\`ege, Belgium \and Royal Observatory of Belgium, Ringlaan 3, 1180 Brussels, Belgium \and Facult\'e Sciences de la Motricit\'e, Universit\'e Libre de Bruxelles, Campus Erasme, Route de Lennik, 808, 1070 Anderlecht, Belgium\\
\email{g.rauw@uliege.be}}
\date{}

  \abstract
      {The long-period O-star binary system HD\,168112 and the triple O-star system HD\,167971 are well-known sources of non-thermal radio emission that arises from a colliding wind interaction. The wind-wind collisions in these systems should result in phase-dependent X-ray emissions. The presence of a population of relativistic electrons in the wind interaction zone could affect the properties of the X-ray emission and make it deviate from the behaviour expected for adiabatic shocks.}{We investigate the X-ray emission of these systems with the goals of quantifying the fraction of the X-ray flux arising from wind interactions and determining whether these emissions follow the predictions for adiabatic wind-wind collisions.}{Six X-ray observations were collected with {\it XMM-Newton}. Three observations were scheduled around the most recent periastron passage of HD\,168112. Spectra and light curves were analysed and compared with simple predictions of model calculations for X-ray emission from colliding wind systems.}{The X-ray emission of HD\,168112 varies as the inverse of the orbital separation, as expected for an adiabatic wind interaction zone. The relative contribution of intrinsic X-ray emission from wind-embedded shocks varies between 38\% at periastron to 81\% at apastron. The wind-wind collision zone remains adiabatic even around periastron passage. The X-ray emission of HD\,167971 displays variations on the orbital timescale of the inner eclipsing binary. The existing data of this system do not allow us to probe variations on the timescale of the outer orbit.}{Shock modification due to the action of relativistic electrons does not seem to be efficiently operating in the HD\,168112 system. In the existing observations, a significant part of the emission of HD\,167971 must arise in the inner eclipsing binary. The origin of this emission is as yet unclear.} 
\keywords{stars: early-type -- stars: individual (HD\,168112) -- stars: individual (HD\,167971) -- binaries: close -- X-rays: stars -- Radio continuum: stars}
\maketitle

\section{Introduction \label{intro}}
Though massive OB stars are rare objects, they are nonetheless key players in many processes in the evolution of galaxies and the Universe as a whole. One of their most remarkable properties is the existence of a powerful radiation-driven stellar wind that combines huge mass-loss rates with highly supersonic outflow velocities \citep[e.g.][]{Vin22}. Most massive stars are found in binary or higher-multiplicity systems \citep[e.g.][]{SF}. 

In such systems, the stellar winds of the components interact, and this interaction can impact the observational properties of these systems over a broad range of wavelengths, from X-rays into the radio domain.
The head-on collision of the highly supersonic winds leads to the formation of an interaction region contained between a pair of oppositely faced, strong hydrodynamic shocks \citep{SBP}. At the shock fronts, the kinetic energy normal to the shock is converted into heat, which may result in post-shock plasma temperatures of $\sim 10$\,MK. This makes colliding wind binaries interesting targets for observations in the X-ray domain \citep[for reviews see][]{RaNa16,Handbook}. However, observations of large samples of massive stars indicate that very strong X-ray emission is only observed for a subset of the massive binaries \citep[e.g.][]{Naze09}.

Another observational signature of wind interactions that is relevant in the context of our present work is synchrotron non-thermal radio emission. Whilst thermal radio emission is naturally expected due to free-free emission from the stellar wind \citep{Wri75}, about 15 -- 25\% of the massive stars in a distance-limited sample were instead found to display a clearly non-thermal radio spectrum \citep{Abb84,Bie89}. This synchrotron emission reveals the presence of a population of relativistic electrons inside the stellar winds. Theoretical models of synchrotron emission from single O-type stars were unable to explain the observed non-thermal radio emission \citep{vLo05}. Intense optical spectroscopic monitoring or interferometric surveys of these non-thermal radio emitter massive stars unveiled the binary or higher-multiplicity nature of these objects \citep[e.g.][]{Dou00,Naz10,San14,9Sgr}. In colliding wind binaries, relativistic electrons are accelerated via the diffusive shock acceleration process at the strong hydrodynamic shocks between the stellar winds of the binary components \citep[e.g.][and references therein]{Rei14,Pit21}. The young open cluster NGC\,6604 \citep{Rei08} hosts two O-type systems (HD\,167971 and HD\,168112) that are prominent non-thermal radio emitters and display a relatively strong X-ray emission \citep{Bie89,FdP04,Blo05,FdP05,Blo07}.

HD\,167971 (MY\,Ser) is a triple system consisting of a close eclipsing inner binary \citep{Lei87,Dav88,May10,Iba13} with an orbital period of 3.32\,days \citep[][and references therein]{Iba13}. The presence of a third component has been known for quite some time \citep{Lei87,Dav88}. \citet{Mai19} classified the components of the eclipsing binary as O4/5\,If + O4/5\,V-III and assigned an O8\,Iaf spectral type to the tertiary component. However, the physical link between the inner binary and the third star was unclear at first. \citet{Blo07} analysed the existing radio observations of HD\,167971, finding a timescale of variation of about 20 years, which they suggested is the orbital period of the third component. Multi-epoch interferometry allowed the gravitational link between the third star and the inner binary to be established \citep{FdP12,LeB17}. From the light-time effect affecting the times of minimum light of MY\,Ser, \citet{Iba13} inferred a period of $21.2 \pm 0.7$\,yr and an eccentricity of $0.53 \pm 0.05$ for the outer orbit of the triple system. Based on interferometric measurements, \citet{LeB17} derived a similar period of $7803 \pm 540$\,d (= $21.4 \pm 1.5$\,yr) and an eccentricity of $0.44 \pm 0.02$ for the outer orbit. 

Confirmation of binarity has been slower in the case of HD\,168112. Based on the temporal variations in the radio flux, \citet{Blo05} suggested a likely orbital period of 1.4\,yr. The first direct evidence for binarity came from Very Large Telescope Interferometer (VLTI) interferometry: \citet{San14} detected a near-equal brightness companion ($\Delta m_H =0.17$\,mag) at a separation of 3.33\,mas (corresponding to a physical separation of 6.7\,AU for a distance of 2\,kpc). Based on partially de-blended lines, \citet{Mai19} assigned the spectral type O4.5\,III(f) to the primary and O5.5\,IV((f)) to the secondary\footnote{These classifications were revised to O4.5\,IV((f)) + O5.5\,V(n)((f)) by \citet{Put23}.}. Quite recently, two groups independently inferred full consolidated orbital solutions for HD\,168112 \citep{Put23,Blo24}. These radial velocity solutions yield a 514\,d (= 1.4\,yr) orbital period and reveal two nearly equal-mass stars in a highly eccentric orbit ($e \sim 0.74 - 0.75$). According to these orbital solutions, HD\,168112 went through its periastron in March 2023. We monitored this event by means of three coordinated {\it XMM-Newton} and \textit{Karl G.\ Jansky} Very Large Array (VLA) observations in the X-ray and radio domains, respectively. In the present paper, we analyse the X-ray data in light of our current knowledge of HD\,168112 and HD\,167971. The new radio observations of HD\,168112 are discussed in detail in \citet{Blo24}. 

\begin{table*}
  \caption{{\it XMM-Newton} observations of HD\,168112 and HD\,167971.}
  \begin{center}
  \begin{tabular}{c c c c c c c c c c}
    \hline
    Obs. & Obs.\ ID. & Duration & JD-2\,450\,000 & EPIC filter & Target &  \multicolumn{2}{c}{$\phi_{\rm HD\,168112}$} & \multicolumn{2}{c}{HD\,167971}\\
    & & (ks) & & & on axis & Putkuri & Blomme & $\phi_{\rm inner}$ & $\phi_{\rm outer}$\\
    \hline
    1 & 0008820301 &  9.1 &  2372.56 & Thick & -- & 0.096 & 0.063 & 0.49 & 0.697\\
    2 & 0008820601 & 13.4 &  2526.78 & Thick & -- & 0.396 & 0.364 & 0.92 & 0.717\\
    3 & 0740990101 & 23.8 &  6909.73 & Medium & HD\,167971 & 0.931 & 0.921 & 0.45 & 0.278\\
    4 & 0920040401 & 12.5 & 10009.81 & Thick & HD\,168112 & 0.968 & 0.974 & 0.75 & 0.676\\
    5 & 0920040501 & 11.5 & 10025.98 & Thick & HD\,168112 & 0.000 & 0.006 & 0.62 & 0.678\\
    6 & 0920040601 & 11.5 & 10042.45 & Thick & HD\,168112 & 0.032 & 0.038 & 0.58 & 0.680\\
    \hline
  \end{tabular}
  \end{center}
  \tablefoot{The durations indicated in the third column correspond to the effective exposure time of the EPIC-MOS camera (i.e.\ after removing the time interval affected by a flare in the first and third observation). The Julian day given in column 4 is evaluated at mid-exposure. The last four columns yield the orbital phases respectively for HD\,168112 (\citealt{Put23} and \citealt{Blo24}), the eclipsing inner binary of HD\,167971 \citep{Iba13} and the orbit of the outer component of HD\,167971 \citep{LeB17}.\label{tab:obs}}
\end{table*}

\section{Observations and data processing \label{sect:obs}}
\subsection{X-ray observations}
Our targets were observed six times with the {\it XMM-Newton} satellite \citep{Jansen}, which carries three mirror modules that focus X-rays onto three European Photon Imaging Cameras \citep[EPIC;][]{MOS,pn} and two Reflection Grating Spectrometer \citep[RGS;][]{RGS} instruments (RGS1 and RGS2). Two of the EPIC instruments use Metal Oxide Semi-conductor \citep[MOS;][]{MOS} charge-coupled device (CCD) arrays whilst the third camera uses a p-n junction CCD \citep[pn;][]{pn}.

Under favourable spacecraft orientations, HD\,168112 and HD\,167971 fall into the same field of view of at least two EPIC cameras. This was the case for all our observations. However, RGS spectra are only available for the source located on-axis. During the first two observations, none of the stars was positioned on-axis. During the third observation, HD\,167971 was observed on-axis, whilst HD\,168112 was positioned on-axis for the three most recent observations. All EPIC observations were taken with the cameras operating in full frame mode; the thick filter was used to reject optical and UV radiation from the targets except for observation 3, for which the medium filter was used. Table\,\ref{tab:obs} provides a journal of the {\it XMM-Newton} observations. The scheduling of the three most-recent observations was optimised to sample orbital phases around HD\,168112's periastron passage in March 2023.   

We processed the data with the Science Analysis System (SAS) software version 19.1.0 using the current calibration files available in January 2024. The first and third observations were affected by soft proton background flares and we discarded the corresponding time intervals. EPIC spectra were extracted from circular regions of radius 30\,\arcsec\ centred on the {\it Gaia} coordinates of the two stars. The background was evaluated over an annulus of inner radius 30\,\arcsec\ and outer radius 50\,\arcsec. For observation 3, we further extracted first and second order RGS1 and RGS2 spectra of HD\,167971. For the last three observations, we did the same for HD\,168112.  

Finally, we built background-corrected light curves of both targets over the 0.5\,keV -- 10.0\,keV band for each of the available EPIC instruments and for each observation. For this purpose, we adopted temporal bin sizes of 100, 500 and 1000\,s. The light curves were corrected to get full point spread function equivalent on-axis count rates using the {\tt epiclccor} SAS task. 

Beside the {\it XMM-Newton} data, we also analysed an 8.7\,ks ROSAT observation (Obs.\ ID\ rp500298n00) taken with the Position Sensitive Proportional Counter (PSPC) instrument between 13 and 15 September 1993. The data were retrieved from the High Energy Astrophysics Science Archive Research Center (HEASARC). We processed these data using the {\tt xselect} software (version 2.4c) and extracted background-corrected spectra for both targets. The source spectrum was obtained from a circular region of 50\arcsec\ radius, whilst the background was extracted from an annulus with inner and outer radii of respectively 50\arcsec\ and 80\arcsec. The HEASARC archive further quotes High Resolution Imager (HRI) count rates for both targets obtained during another 36.0\,ksec ROSAT observation (Obs.\ ID\ rh201995n00) performed between 12 September 1995 and 9 October 1995.

\subsection{Radio data}
Three radio observations of HD\,168112 were obtained around the March 2023 periastron passage using the VLA at the National Radio Astronomy Observatory (NRAO\footnote{The NRAO is a facility of the US National Science Foundation operated under cooperative agreement by Associated Universities, Inc.}). These observations are part of our coordinated {\it XMM-Newton} and VLA monitoring. Data were acquired on three dates (3 March, 18 March, and 5 April 2023), with each date within a few days of the {\it XMM-Newton} observations 4 to 6 listed in Table\,\ref{tab:obs}. They correspond to phases 0.969, 0.998, and 0.033 using the \citet{Blo24} ephemerides. Each VLA observation covers the X (3.6\,cm) and C (6\,cm) bands, with an on-target duration of $\sim14$\,minutes for each band. More details about these radio observations and their data processing are presented in \citet{Blo24}.

\section{X-ray data analysis}
\subsection{Light curves}
We performed $\chi^2$ tests to assess the significance of short-term variability against the hypothesis of a constant count rate. We further tested the significance of several possible trends: a linear time dependence, or a quadratic trend with time. Testing the null hypothesis of a constant count rate, we found that the majority of the light curves do not display any significant variability, that is to say, the significance level $SL$ of the null hypothesis remains above 5\%. A handful of light curves with time bins of 100\,s have $SL < 5$\%, but their 1\,ks counterparts (for the same instrument during the same observation) always have $SL >> 5$\%. Furthermore, the apparent detection of significant variations concerns only a single instrument. Indeed, light curves with 100\,s time bins from the other two EPIC instruments during the same observation have $SL > 5$\%, showing that these apparent detections are statistical flukes. A few light curves are better represented by a linear or quadratic trend, but again this is not confirmed by light curves recorded by the other EPIC instruments during the same observation. We thus conclude that neither of the two binaries displays significant intra-pointing variations beyond the fluctuations consistent with Poisson noise.

\subsection{Spectral fits \label{sect:spec}}
To characterise the spectral energy distributions of both targets in the X-ray domain, we adjusted their {\it XMM-Newton} spectra using version 12.9.0i of the {\tt xspec} code \citep{Arnaud}. For each observation, all available spectra (EPIC and RGS) of a given target were fitted simultaneously. To avoid biases due to the fact that the RGS spectra exist only for a subset of the observations, we also performed fittings of the sole EPIC data for those observations where RGS data were available. We found that the EPIC only and EPIC+RGS spectral fits were fully consistent with each other and all model parameters overlapped well within their error bars.  

High-resolution X-ray spectra of O-type stars unveil a wealth of emission lines of highly ionised species. These spectra indicate that the X-ray emission arises from an optically thin thermal plasma. We thus tested models of the kind 
\begin{equation}
  {\tt TBabs}*{\tt phabs}*\sum_{i=1}^N {\tt apec}(kT_i).
  \label{fitexpression}
\end{equation}
The {\tt apec} components yield the X-ray emission of collisionally ionised, optically thin thermal plasma components \citep{apec}. In our models we used $N=2$ and $N=3$ as we found that a single temperature plasma model ($N=1$) was unable to provide a good fit over the full energy range of the data. 
The {\tt TBabs} model \citep{Wilms} accounts for the photoelectric absorption by the interstellar medium (ISM) along our sightline towards the source. For HD\,167971, \citet{ism} quoted a total interstellar neutral hydrogen column density of $(5.4 \pm 4.0) \times 10^{21}$\,cm$^{-2}$. Whilst the error bar on this column density is large, we note that the value is fully consistent with the mean relation between $N($H\,{\sc i}$)$ and $E(B-V)$. Indeed, for $E(B-V) = 0.87$, the \citet{Boh78} relation yields $5.0 \times 10^{21}$\,cm$^{-2}$, whilst the relation of \citet{ism} yields $5.3\times 10^{21}$\,cm$^{-2}$. In our models, we thus fixed the ISM column of HD\,167971 to $5.15\times 10^{21}$\,cm$^{-2}$, which is the mean of the values obtained with the two $N($H\,{\sc i}$)$ -- $E(B-V)$ relations. For HD\,168112, no direct measurements of the neutral hydrogen column are available. For $E(B-V) = 0.97$ the same relations yield columns of $5.6 \times 10^{21}$\,cm$^{-2}$ and $5.9\times 10^{21}$\,cm$^{-2}$. In our fits, we therefore fixed the ISM column of HD\,168112 to $5.75\times 10^{21}$\,cm$^{-2}$. Our spectral models further included a {\tt phabs} multiplicative absorption component to account for additional photoelectric absorption within the stellar winds, with the corresponding hydrogen column density considered a free parameter of the model.
We stress here that the main objective of our spectral fits is to obtain a good description of the overall X-ray spectral energy distribution. Whilst the model parameters likely reflect the mean properties of the plasma, they should not be over-interpreted, especially because of the model degeneracies between the action of absorbing material and the intrinsic hardness of the emitting plasma. This remark is especially relevant for the lower temperature plasma component and explains the large error bars on the normalisation of the softest {\tt apec} component. However, since this component contributes only a small fraction of the overall flux, this large error has little impact on the flux derivation.

\subsubsection{HD\,168112 \label{fits168112}}
\begin{figure}[htb]
\begin{center}
\resizebox{8.5cm}{!}{\includegraphics{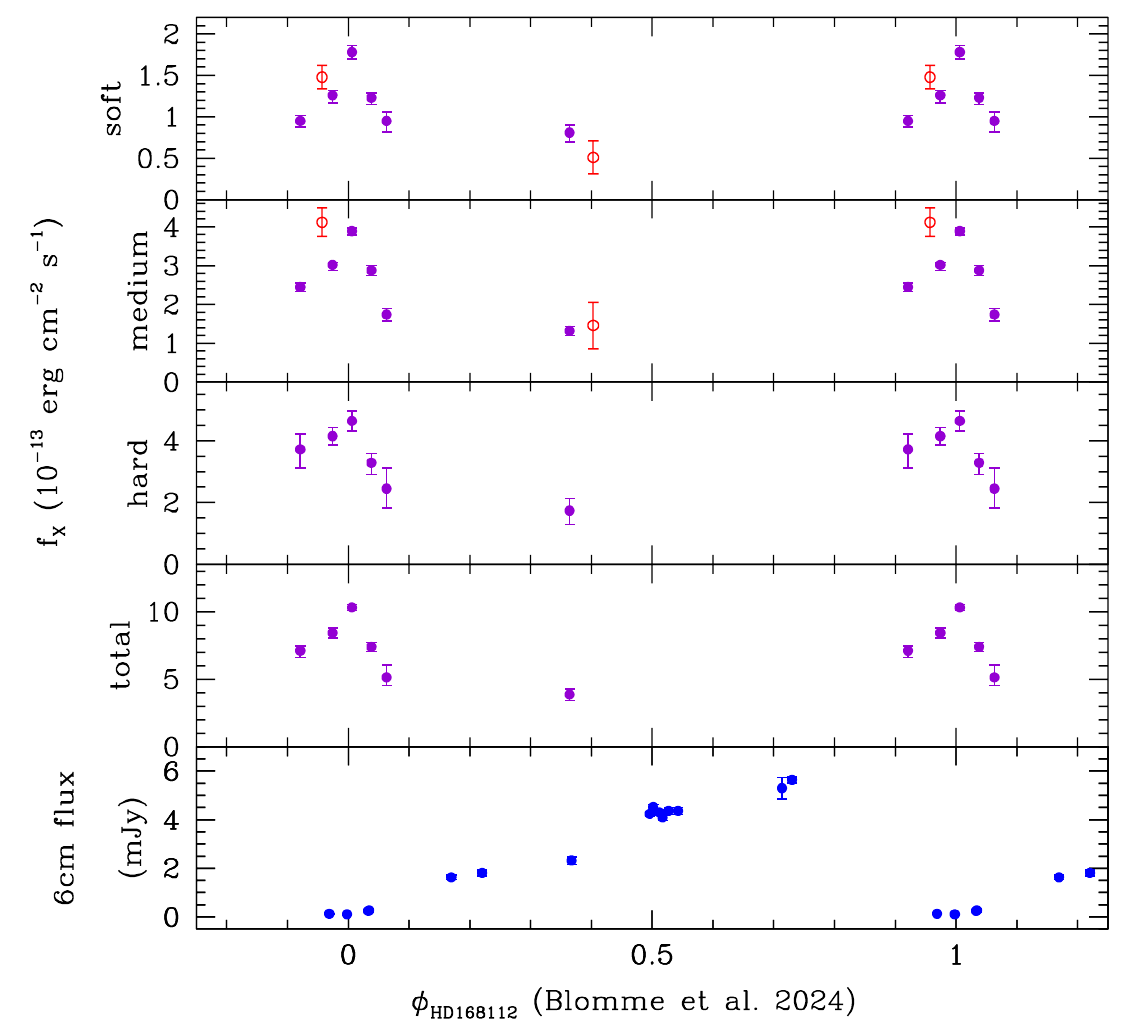}}
\end{center}
\caption{Observed X-ray and radio flux of HD\,168112 as a function of the orbital phase evaluated with the ephemerides of \citet{Blo24}. The top four panels correspond to the different energy bands in the X-ray domain: soft (0.5\,keV -- 1.0\,keV), medium (1.0\,keV -- 2.0\,keV), hard (2.0\,keV -- 4.0\,keV), and total (0.5\,keV -- 10\,keV). Violet filled symbols correspond to {\it XMM-Newton} observations, and the red open circles indicate the ROSAT data. The error bars on the X-ray fluxes correspond to the 90\% confidence intervals. The bottom panel presents the 6\,cm radio flux data from \citet{Blo05}. \label{variab168112}}
\end{figure}

The X-ray spectra of HD\,168112 can be reasonably well described by a two-temperature (2-T) plasma model. The above-mentioned degeneracy between wind column density and plasma temperature is especially apparent when comparing the parameters of observation 2 with those of the other exposures. Adding a third plasma component significantly improves the quality of the fits for the most recent exposures that have the best statistics (see Table\,\ref{tab:fit3THD168112}). The improvement is marginal for observations 1 and 2, which have the poorest statistics. Since the 3-T fits yield a better adjustment of the overall spectral energy distribution, we use the flux values inferred from those fits in what follows. The total X-ray flux of HD\,168112 exhibits highly significant variations between the various observations (see Table\,\ref{tab:fit3THD168112} and Fig.\,\ref{variab168112}). This result is independent of the model (2-T or 3-T), although, as indicated before, the 3-T models better represent the true flux of the source.

Owing to the smaller energy range covered by the ROSAT-PSPC instrument, the PSPC spectrum was fitted using a 1-T model. Assuming the same set of 1-T model parameters (except for the normalisation) describes the properties of the source during the ROSAT-HRI observation, we used the HEASARC WebPIMMS tool\footnote{http://heasarc.gsfc.nasa.gov/cgi-bin/Tools/w3pimms/w3pimms.pl} to convert the HRI count rates into fluxes in the 0.5\,keV -- 1.0\,keV and 1.0\,keV -- 2.0\,keV energy domains.  

As outlined in Sect.\,\ref{intro}, HD\,168112 and HD\,167971 are known as synchrotron non-thermal radio emitters. This synchrotron emission reflects the presence of a population of relativistic electrons accelerated by diffusive shock acceleration in the colliding wind region. In principle, inverse Compton (IC) scattering of stellar UV photons by those relativistic electrons could result in a non-thermal X-ray emission with a power law photon index $\Gamma = \frac{n+1}{2}$ where $n$ is the index of the energy distribution of the relativistic electrons \citep{CW91}. To check for the existence of such a non-thermal X-ray emission, we adjusted models of the kind  
\begin{equation}
  {\tt TBabs}*{\tt phabs}*\left(\sum_{i=1}^2 {\tt apec}(kT_i) + {\tt power}\right),
  \label{fitexpression_power}
\end{equation}
where the {\tt power} component stands for a non-thermal emission described by a power law relation. This model usually resulted in fits of very similar quality as for the 3-T thermal plasma model. The best-fit photon index of the power law turned out to be highly variable between exposures, changing from 2.2 to 6.9. These indices differ significantly from the value expected for a population of relativistic electrons ($\Gamma \sim 1.5$) if the shocks responsible for the acceleration have the usual compression ratio $\sim 4$. Moreover, their erratic variations further indicate that there currently is no clear evidence for the existence of a non-thermal component in the X-ray emission of HD\,168112. Moreover, the best quality EPIC-pn spectra (observations 4, 5 and 6) display an Fe\,{\sc xxv} line at 6.7\,keV, which is a clear indication that the harder X-ray emission arises inside a very hot thermal plasma. We come back to this point in Sect.\,\ref{CWB168112}. 

\begin{table*}[htb]
  \caption{Fits of the X-ray spectra of HD\,168112 using 3-T plasma models.\label{tab:fit3THD168112}}
  \tiny
  \begin{center}
  \begin{tabular}{c c c c c c c c c c c c}
    \hline
    Obs. & Instruments & $N_{\rm H}$ & $kT_1$ & norm$_1$ & $kT_2$ & norm$_2$ & $kT_3$ & norm$_3$ & $\chi^2_{\nu}$ & d.o.f. & $f_X$ (0.5 - 10\,keV) \\
    \vspace*{-2mm}\\
    & & ($10^{22}$\,cm$^{-2}$) & (keV) & (cm$^{-5}$) & (keV) & (cm$^{-5}$) & (keV) & (cm$^{-5}$) & & & ($10^{-14}$\,erg\,cm$^{-2}$\,s$^{-1}$) \\
    \hline
    \vspace*{-2mm}\\
    1 & MOS1+2 & $0.54^{+0.18}_{-0.20}$ & $0.20^{+.13}_{-.15}$ & $\left(4.4^{+157.9}_{-3.6}\right)\,10^{-3}$ & $0.56^{+.22}_{-.19}$ & $\left(7.3^{+4.0}_{-6.1}\right)\,10^{-4}$ & $\geq 2.8$ & $\left(2.3^{+1.2}_{-0.8}\right)\,10^{-4}$ & 0.94 & 66 & $51.4^{+9.4}_{-5.9}$ \\
    \vspace*{-2mm}\\
    2 & MOS1+2 & $0.26^{+0.20}_{-0.20}$ & $0.14^{+.83}_{-.14}$ & $\left(2.7^{+297.5}_{-2.6}\right)\,10^{-3}$ & $0.60^{+.08}_{-.12}$ & $\left(4.5^{+2.2}_{-1.1}\right)\,10^{-4}$ & $\geq 2.5$ & $\left(1.9^{+0.8}_{-0.7}\right)\,10^{-4}$ & 0.76 & 61 & $38.7^{+2.7}_{-4.5}$ \\
    \vspace*{-2mm}\\
    3 & EPIC & $0.59^{+0.13}_{-0.27}$ & $0.26^{+.05}_{-.03}$ & $\left(4.4^{+4.7}_{-3.5}\right)\,10^{-3}$ & $1.26^{+.20}_{-.38}$ & $\left(5.1^{+1.3}_{-2.4}\right)\,10^{-4}$ & $\geq 2.7$ & $\left(2.5^{+2.8}_{-0.8}\right)\,10^{-4}$ & 0.85 & 162 & $71.2 \pm 4.9$ \\
    \vspace*{-2mm}\\
    4 & EPIC+RGS & $0.28^{+0.20}_{-0.20}$ & $0.33^{+.12}_{-.05}$ & $\left(9.4^{+19.4}_{-7.0}\right)\,10^{-4}$ & $0.94^{+.09}_{-.08}$ & $\left(3.9^{+1.1}_{-1.0}\right)\,10^{-4}$ & $2.9^{+0.5}_{-0.4}$ & $\left(5.9^{+0.8}_{-0.9}\right)\,10^{-4}$ & 0.96 & 451 & $84.3 \pm 3.7$ \\    
    \vspace*{-2mm}\\
    5 & EPIC+RGS & $0.33^{+0.11}_{-0.13}$ & $0.31^{+.09}_{-.03}$ & $\left(2.0^{+1.4}_{-1.2}\right)\,10^{-3}$ & $0.98^{+.07}_{-.07}$ & $\left(5.8^{+1.4}_{-1.3}\right)\,10^{-4}$ & $2.9^{+0.6}_{-0.4}$ & $\left(6.3^{+1.2}_{-1.0}\right)\,10^{-4}$& 1.11 & 493 & $103.1 \pm 2.3$ \\
    \vspace*{-2mm}\\
    6 & EPIC+RGS & $0.52^{+0.11}_{-0.11}$ & $0.30^{+.08}_{-.06}$ & $\left(2.2^{+2.7}_{-1.1}\right)\,10^{-3}$ & $0.76^{+0.16}_{-0.07}$ & $\left(6.5^{+3.0}_{-2.2}\right)\,10^{-4}$& $3.2^{+1.1}_{-0.6}$ & $\left(4.3^{+0.9}_{-0.9}\right)\,10^{-4}$ & 1.03 & 414 & $74.0 \pm 3.3$ \\
    \vspace*{-2mm}\\
    \hline
  \end{tabular}
  \end{center}
  \tablefoot{The model was defined according to Eq.\,(\ref{fitexpression}) with $N=3$. The ISM column was fixed to $5.75 \times 10^{21}$\,cm$^{-2}$. Abundances were fixed to solar following \citet{Asplund}. The plasma normalisation parameters are equal to $\frac{10^{-14}}{4\,\pi\,D^2}\,\int n_e\,n_{\rm H}\,dV$ in units cm$^{-5}$ with $D$ the distance of the source and $\int n_e\,n_{\rm H}\,dV$ the emission measure of the plasma. The last column yields the observed flux in the 0.5 -- 10\,keV band. The quoted errors correspond to the 90\% confidence intervals. The uncertainties on the observed fluxes were evaluated using the {\tt cflux} command of the {\tt xspec} software.}
\end{table*}

\subsubsection{HD\,167971 \label{fits167971}}
HD\,167971 is a few times brighter in X-rays than HD\,168112. Its X-ray spectra are again reasonably well described by means of 2-T plasma models. However, including a third plasma component significantly improves the quality of the fits for the majority of the exposures (see Table\,\ref{tab:fit3THD167971}). We observe significant variations in the total flux between the different exposures, although they are of lower amplitude than in the case of HD\,168112 (see Table\,\ref{tab:fit3THD167971}). The highest quality EPIC-pn spectrum (observation 3) displays a clear Fe\,{\sc xxv} emission near 6.7\,keV consistent with a rather high plasma temperature. The ROSAT data were analysed in the same way as for HD\,168112. 

As for HD\,168112, we also adjusted models including a power law component as of Eq.\,\ref{fitexpression_power}. Compared to the 3-T models, the 2-T + power law model resulted in a slight improvement of the quality of the fits in 4 cases out of 6. We found again best-fit photon indices that were quite different from the expected value of 1.5. For all observations, except the third one, $\Gamma$ was found to be in the range 2.6 -- 3.4. For the third observation, the best agreement with the data was achieved for $\Gamma = 7.9$. Again, we conclude that there is currently no evidence for the existence of a genuine non-thermal component in the X-ray emission of HD\,167971. We return to this point in Sect.\,\ref{CWB167971}. 

\begin{table*}[htb]
  \caption{Fits of the X-ray spectra of HD\,167971 using 3-T plasma models. \label{tab:fit3THD167971}}
  \tiny
  \begin{center}
  \begin{tabular}{c c c c c c c c c c c c}
    \hline
    Obs. & Instruments & $N_{\rm H}$ & $kT_1$ & norm$_1$ & $kT_2$ & norm$_2$ & $kT_3$ & norm$_3$ & $\chi^2_{\nu}$ & d.o.f. & $f_X$ (0.5 - 10\,keV) \\
    \vspace*{-2mm}\\
    & & ($10^{22}$\,cm$^{-2}$) & (keV) & (cm$^{-5}$) & (keV) & (cm$^{-5}$) & (keV) & (cm$^{-5}$) & & & ($10^{-13}$\,erg\,cm$^{-2}$\,s$^{-1}$) \\
    \hline
    \vspace*{-2mm}\\
    1 & MOS1+2 & $0.72^{+0.11}_{-0.12}$ & $0.31^{+.07}_{-.03}$ & $\left(11.9^{+7.9}_{-6.0}\right)\,10^{-3}$ & $0.86^{+.12}_{-.15}$ & $\left(12.8^{+7.7}_{-5.5}\right)\,10^{-4}$ & $3.0^{+2.6}_{-0.9}$ & $\left(6.5^{+3.1}_{-2.6}\right)\,10^{-4}$ & 0.85 & 111 & $16.6 \pm 1.1$ \\
    \vspace*{-2mm}\\
    2 & EPIC & $0.64^{+0.06}_{-0.07}$ & $0.30^{+.03}_{-.02}$ & $\left(8.6^{+3.3}_{-2.6}\right)\,10^{-3}$ & $0.98^{+.07}_{-.07}$ & $\left(12.9^{+2.6}_{-2.6}\right)\,10^{-4}$ & $3.1^{+2.8}_{-0.8}$ & $\left(4.4^{+2.1}_{-1.8}\right)\,10^{-4}$ & 1.11 & 231 & $14.1 \pm 0.7$ \\
    \vspace*{-2mm}\\
    3 & EPIC+RGS & $0.60^{+0.04}_{-0.04}$ & $0.30^{+.02}_{-.01}$ & $\left(6.3^{+1.6}_{-1.4}\right)\,10^{-3}$ & $0.81^{+.04}_{-.03}$ & $\left(16.9^{+2.3}_{-2.2}\right)\,10^{-4}$ & $2.0^{+0.2}_{-0.1}$ & $\left(8.4^{+1.0}_{-1.1}\right)\,10^{-4}$ & 1.13 & 967 & $15.8 \pm 0.4$ \\
    \vspace*{-2mm}\\
    4 & EPIC & $0.53^{+0.09}_{-0.09}$ & $0.31^{+.08}_{-.03}$ & $\left(6.6^{+4.4}_{-3.3}\right)\,10^{-3}$ & $0.84^{+.12}_{-.08}$ & $\left(16.6^{+6.0}_{-2.9}\right)\,10^{-4}$ & $1.9^{+0.5}_{-0.3}$ & $\left(10.8^{+2.6}_{-2.7}\right)\,10^{-4}$ & 1.09 & 180 & $19.1 \pm 0.9$ \\    
    \vspace*{-2mm}\\
    5 & EPIC & $0.40^{+0.09}_{-0.15}$ & $0.15^{+.22}_{-.06}$ & $\left(10.2^{+63.6}_{-7.4}\right)\,10^{-3}$ & $0.65^{+.33}_{-.05}$ & $\left(21.0^{+6.1}_{-7.7}\right)\,10^{-4}$ & $1.3^{+0.1}_{-0.1}$ & $\left(19.3^{+2.4}_{-4.6}\right)\,10^{-4}$& 1.22 & 186 & $20.0 \pm 0.9$ \\
    \vspace*{-2mm}\\
    6 & EPIC & $0.47^{+0.16}_{-0.17}$ & $0.34^{+.21}_{-.07}$ & $\left(3.9^{+7.5}_{-2.9}\right)\,10^{-3}$ & $0.79^{+0.13}_{-0.09}$ & $\left(16.3^{+7.2}_{-6.4}\right)\,10^{-4}$& $1.9^{+0.5}_{-0.3}$ & $\left(12.5^{+3.3}_{-3.1}\right)\,10^{-4}$ & 1.15 & 143 & $19.1 \pm 0.9$ \\
    \vspace*{-2mm}\\
    \hline
  \end{tabular}
  \end{center}
  \tablefoot{Same as Table\,\ref{tab:fit3THD168112}, but the ISM column was here fixed to $5.15 \times 10^{21}$\,cm$^{-2}$. For observation 5, a few deviating energy bins of the EPIC-pn spectrum below 0.4\,keV were excluded from the fitting procedure.}
\end{table*}

\section{Discussion \label{discussion}}
The properties of the hot plasma in a colliding wind interaction zone depend strongly on the efficiency of radiative cooling \citep{SBP}. If the plasma density is high, collisional recombinations will be frequent resulting in a fast radiative cooling. As a result, the plasma temperature drops rapidly before the material escapes the system. This situation applies notably to massive binaries with short orbital periods such as the inner eclipsing binary in HD\,167971. On the contrary, in wide massive binaries, such as HD\,168112 or the outer system of HD\,167971, the density of the plasma in the post-shock region of the colliding winds interaction is comparatively low. As a result, radiative cooling should be negligible and the hot shocked plasma only cools adiabatically. Under these circumstances, the X-ray luminosity is expected to scale with $1/r$ where $r$ is the instantaneous orbital separation between the stars \citep{SBP,Pit18}.

The efficiency of radiative cooling is usually quantified through the ratio between the radiative cooling time and the escape time from the shock region $\chi = \frac{t_{\rm cool}}{t_{\rm esc}} = \frac{v_{\infty}^4\,r}{\dot{M}}$ \citep{SBP}. Here $r$ stands for the instantaneous orbital separation of the stars, $\dot{M}$ and $v_{\infty}$ are the wind mass-loss rate and wind terminal velocity (assuming the wind has the time to accelerate to its terminal velocity before the collision). Adopting the spectral types inferred by \citet{Mai19} or those quoted by \citet{Put23} along with the corresponding mass-loss rates and wind velocities from \citet{Muijres}, we infer $\chi \geq 22$ for both shocked winds of HD\,168112. This result holds at all orbital phases, with the lowest value being reached at periastron. Hence, the radiative cooling via collisional recombination should be negligible at all orbital phases.

Collisional recombination might not be the sole cooling mechanism affecting the post-shock plasma. \citet{Mac23} drew attention to the importance of IC cooling that can become the primary cooling mechanism notably in eccentric binary systems with large wind momentum ratios. IC cooling of the thermal electrons could make the shocked winds strongly radiative near periastron despite the fact that  the \citet{SBP} criterion would indicate them to be in the adiabatic regime. \citet{Mac23} therefore formulate a specific criterion to evaluate the importance of IC cooling: $\chi_{\rm IC} = \frac{1.61\,x^{\rm stag}_{12}\,v_8}{L_5}$. Here $x^{\rm stag}_{12}$ stands for the distance from the centre of the star to the shock in units $10^{12}$\,cm, $v_8$ is the pre-shock velocity in units 1000\,km\,s$^{-1}$ and $L_5$ the bolometric luminosity of the star in units $10^5\,L_{\odot}$. If $\chi_{\rm IC} < 1$, then IC cooling makes the shock radiative. Applying this criterion to HD\,168112 at periastron, we infer $\chi_{\rm IC} \geq 8$ for both components of the system. Hence, we conclude that both criteria suggest that the wind interaction zone remains in the adiabatic regime all around the orbital cycle.

Applying the same calculations to the inner binary of HD\,167971 yields $\chi = 3.2\,(v/v_{\infty})^4$ and $\chi_{\rm IC} = 0.9\,(v/v_{\infty})$. Because of the proximity of the two stars and their mutual radiation fields, their winds do not accelerate to $v_{\infty}$ before they collide. Assuming that they reach $v_{\infty}/2$, which seems rather optimistic, we would find $\chi \simeq 0.2$ and $\chi_{\rm IC} \simeq 0.4$, indicating that the shocked winds should be in the radiative regime. For the wind interaction between the inner binary and the tertiary component, the cooling parameters vary with phase because of the changing orbital separation. In this case, we find $\chi \geq 295$ and $\chi_{\rm IC} \geq 75$.

Wide eccentric binaries offer an ideal testbed for the theory of adiabatic wind interactions, and more specifically for the predicted $1/r$ dependence of the X-ray flux. Evidence for such a behaviour was found in several O + O and Wolf-Rayet (WR) + O binaries \citep[e.g.][]{Naz12,Pan14,Gos16}, but strong deviations from this simple expectation were observed in other systems among which the long-period (8 -- 9\,years) binaries 9\,Sgr \citep{9Sgr} and WR\,140 \citep{Zhe21}. At first sight, some departures from the $1/r$ scaling could be due to radiative inhibition or braking \citep{SP94,Gay97}. However, the efficiency of this mechanism is strongly reduced by the wide separation between the stars even around periastron passage. For 9\,Sgr and WR\,140, part of the explanation may reside instead in the fact that both systems are non-thermal radio sources and, thus, host relativistic electrons in their wind interaction zone. Indeed, the pressure that relativistic particles, accelerated by the diffusive shock acceleration mechanism, exert on the pre-shock flow should indeed lead to shock modification \citep{Pit06}. Both systems investigated in the present study display non-thermal radio emission indicating the presence of populations of relativistic electrons. Both systems feature a wind-wind collision that arises in a highly eccentric orbit with a wide orbital separation. They are thus ideal targets to investigate the impact of shock modification on the X-ray emission of colliding wind systems.

\begin{figure}
  \begin{center}
    \resizebox{8cm}{!}{\includegraphics{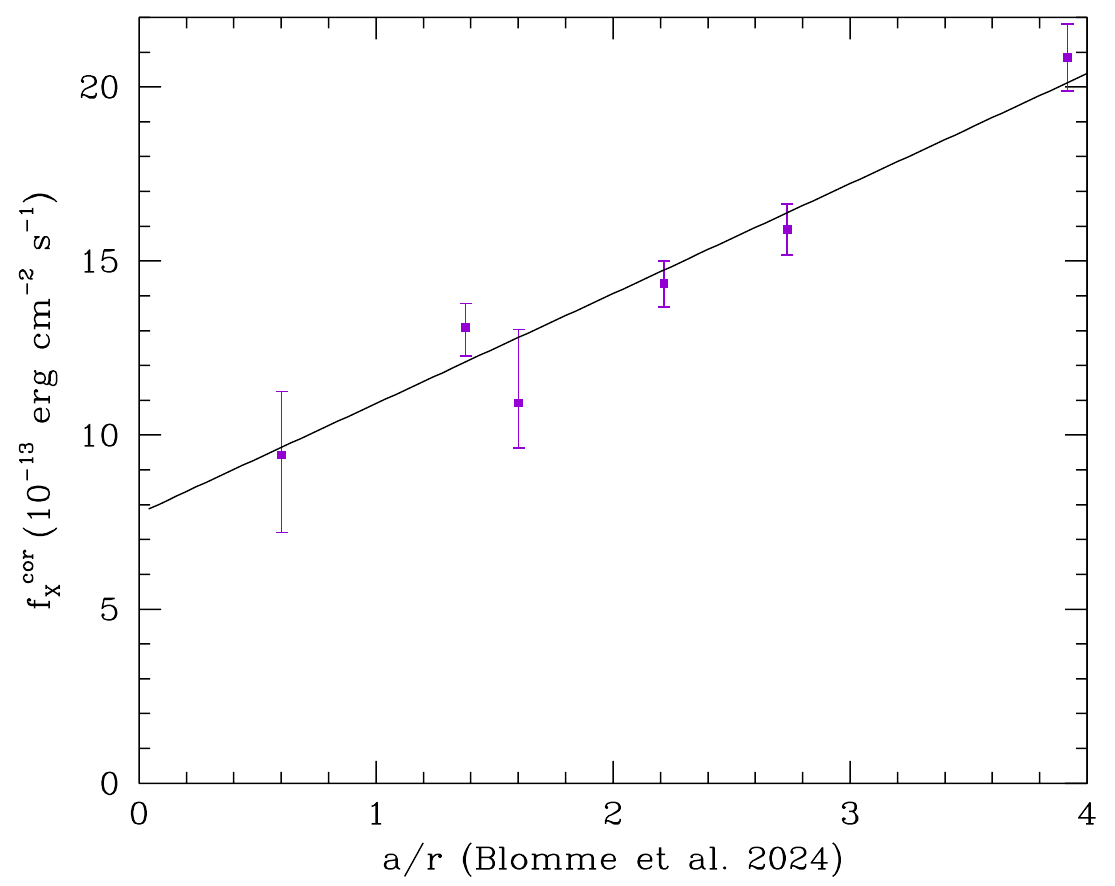}}
  \end{center}
  \caption{X-ray flux of HD\,168112 in the 0.5\,keV -- 10\,keV energy band corrected for absorption by the ISM as a function of $a/r$ computed according to the orbital solution from \citet{Blo24}. The straight line corresponds to the linear fit given by Eq.\,\ref{eq1d}.\label{1surD}} 
  \begin{center}
    \resizebox{8cm}{!}{\includegraphics{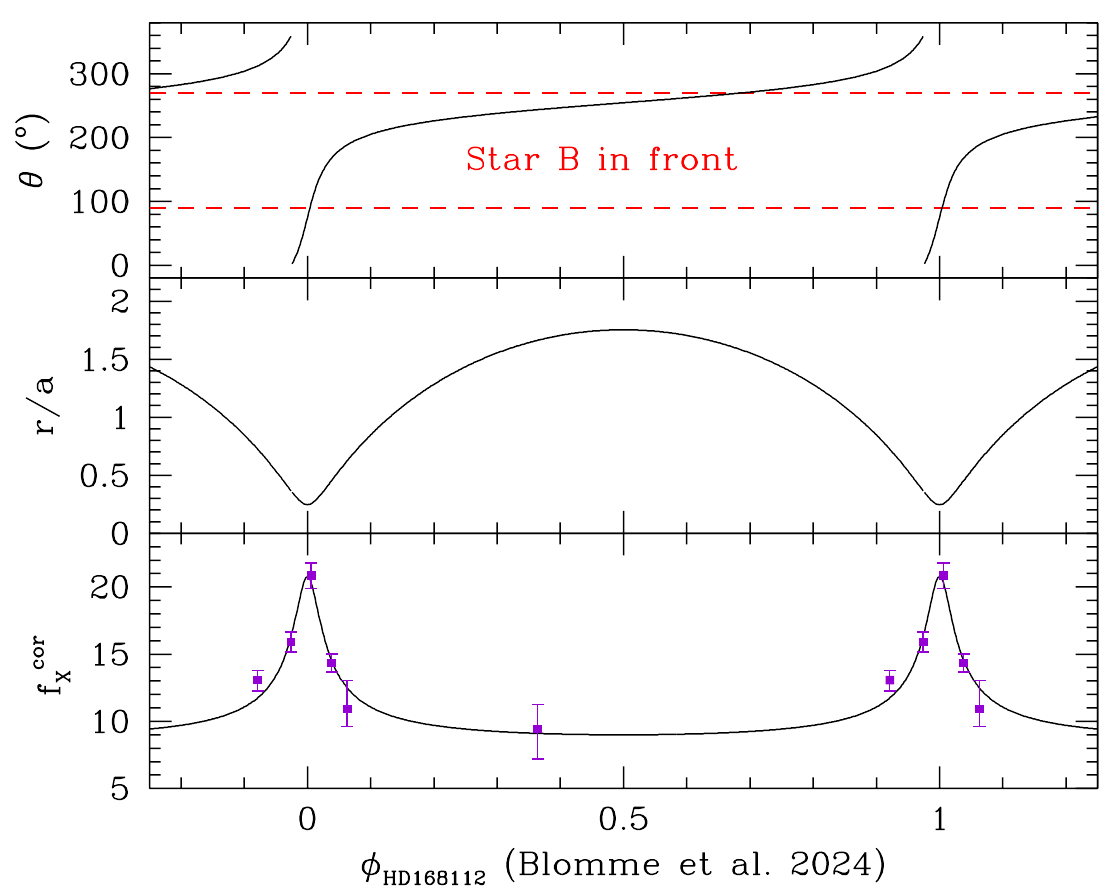}}
  \end{center}
  \caption{Variation in the position angle (top panel) and the orbital separation ($r/a$) as a function of the orbital phase computed according to the orbital solution from \citet{Blo24}. The position angle is measured in the orbital plane and is defined here as the angle between the line joining the two stars and the direction of conjunction with the primary star (labelled component A in \citealt{Blo24}) in front. Phase 0.0 corresponds to periastron passage. Bottom panel: X-ray flux of HD\,168112 in the 0.5\,keV -- 10\,keV energy band corrected for absorption by the ISM as a function of orbital phase along with the best fit $1/r$ scaling of the colliding wind X-ray emission given by Eq.\ref{eq1d}.\label{orbit}}
\end{figure}

\subsection{Colliding winds in HD\,168112 \label{CWB168112}}
Figure\,\ref{variab168112} illustrates the variability of the X-ray and radio fluxes of HD\,168112 with orbital phase computed using the ephemerides of \citet{Blo24}. Quite similar results are obtained if we use the ephemerides of \citet{Put23}, although there is a small shift in phase\footnote{We focus on the results obtained with the \citet{Blo24} orbital solution as it contains significantly more data points around periastron passage and should thus result in better constrained orbital elements.} (see Table\,\ref{tab:obs}). The X-ray flux displays a significant increase, by a factor of 2 to 3, around periastron passage. Whilst this increase is observed in all energy bands, the strongest variations (by a factor of 3.0) occur in the medium energy band, whilst the soft energy band displays the lowest increase (by a factor of 2.2).

Figure\,\ref{1surD} illustrates the X-ray flux in the 0.5 -- 10\,keV energy band, corrected for absorption by the ISM, as a function of the ratio $a/r$ computed from the orbital solution of \citet{Blo24}. Here $a$ stands for the orbital semi-major axis, whilst $r$ is the instantaneous orbital separation. The straight line yields a least square fit to these data given by
\begin{equation}
  f_{\rm X} (10^{-13}\,{\rm erg}\,{\rm cm}^{-2}\,{\rm s}^{-1}) = (3.16 \pm 0.42)\,(a/r) + (7.75 \pm 1.05)
  \label{eq1d}
.\end{equation}
Using instead the \citet{Put23} orbital solution yields a very similar relation with the numerical values of $3.30 \pm 0.45$ and $7.41 \pm 1.10$. 
The best-fit relation of Eq.\,\ref{eq1d} is illustrated in Fig.\,\ref{orbit} along with the phase dependence of $r/a$ and of the position angle $\theta$ defined as the true anomaly variation between the conjunction with the O4.5\,IV primary star in front and the actual direction of the line joining the two stars. We observe no obvious dependence of the X-ray fluxes on $\theta$, which indicates that the observed variations are not due to absorption effects.   

In shorter period systems (e.g.\ WR~21a, P$_{\rm orb} = 31.7$\,days and Cyg~OB2 \#8a, P$_{\rm orb} = 21.9$\,days), the X-ray emission displays an asymmetric behaviour between phases before and after periastron \citep[e.g.][]{Gos16,Mos20}. This was attributed to disruptions of the shock at periastron, and the shock subsequently needing some time to recover as the orbital separation increased again. Our observations of HD\,168112 at $\phi = 0.974$ (observation\,4) and $\phi = 0.038$ (observation\,6) allowed us to check for the existence of such a hysteresis-like behaviour. Table\,\ref{tab:fit3THD168112} clearly shows the similarity in properties of these two phases. Therefore, we find no significant evidence for such a phenomenon in the HD\,168112 system: the variations in the X-ray flux appear rather symmetrical around periastron passage.

As described above, the presence of relativistic electrons in the wind interaction zone of HD\,168112 could result in shock modification \citep{Pit06}. This should in turn lower the X-ray luminosity (compared to the $1/r$ relation) and the post-shock temperature and, hence, the temperature of the X-ray emitting plasma \citep{Pit06}. Figures\,\ref{1surD} and \ref{orbit} do not show any evidence for a significant deviation of the X-ray flux from the $1/r$ relation. 

As outlined in Sect.\,\ref{intro}, HD\,168112 displays a non-thermal radio emission that is most likely associated with the wind interaction zone. \citet{FdP24} presented European Very Long Baseline Interferometry Network (EVN) radio observations of HD\,168112 collected in November 2019 with an angular resolution of $12 \times 22$\,mas$^2$. The date of the EVN observations corresponds to phase $\phi = 0.598$ of HD\,168112 according to the \citet{Blo24} ephemerides. The radio emission of HD\,168112 was partially resolved and found to be elongated, consistent with the emission expected from a colliding wind interaction.

The bottom panel of Fig.\,\ref{variab168112} illustrates the variations in the observed radio emission at 6\,cm wavelength \citep[using our new VLA radio observations along with the radio light curve of][]{Blo05}. This radio emission displays a pronounced minimum around periastron passage. This minimum could reflect a genuine disruption of the shock around periastron passage. Indeed, if the shock were disrupted (as seen in WR\,21a or Cyg\,OB2 \#8a), the particle acceleration mechanism would be temporarily switched off, resulting in a strong attenuation of the synchrotron radio emission. However, the symmetrical variations in the X-ray flux around periastron clearly indicate that the shock is not disrupted. Instead, the most likely explanation of the radio minimum is the substantial free-free absorption by the stellar wind material. Indeed, as the stars approach periastron, the colliding wind interaction moves deeper into the optically thick winds and the synchrotron radio emission is strongly attenuated \citep{Blo05}. From the variations in the position angle in Fig.\,\ref{orbit}, we also see that the wind interaction zone remains partially hidden behind the radio photosphere of the secondary until about phase 0.35 and fully emerges only once the orbital separation increases as the stars approach apastron. This situation is quite reminiscent of the behaviour observed for Cyg\,OB2 \#9 \citep[O5-5.5\,I + O3-4\,III, P$_{\rm orb} = 858.4$\,d, $e = 0.71$,][]{Naz12,Blo13}. We refer to \citet{Blo24} for a more extensive discussion of the radio emission of HD\,168112.

The VLTI observation analysed by \citet{San14} corresponds to orbital phase $\phi = 0.334$ with the \citet{Blo24} ephemerides. At this phase, the orbital separation amounts to $1.62\,a$, and $\theta$ is equal to $241^{\circ}$. The third {\it Gaia} data release yields a parallax of $0.498 \pm 0.020$\,mas for HD\,168112. Together with the $a\,\sin{i} = 897$\,R$_{\odot}$, the angular separation yields an inclination of $i \simeq 63^{\circ}$. Combining this inclination with the minimum masses inferred by \citet{Blo24}, we obtain absolute masses for the two stars of 26.1 and 25.9\,M$_{\odot}$. Using instead the orbital solution of \citet{Put23}, that is, $a\,\sin{i} = 1005$\,R$_{\odot}$, we obtain $i \simeq 77^{\circ}$ and absolute masses of the two stars of 29.2 and 26.5\,M$_{\odot}$. The components of HD\,168112 were classified as O4.5\,IV((f)) for the primary and O5.5\,V(n)((f)) for the secondary \citep{Put23}. In both cases, the absolute masses that we derive from the comparison of the orbital solutions with the interferometric measurement are lower by $\sim 25$\% than the masses expected for stars of these spectral types \citep{Martins}.

The bolometric luminosities of the components were given as $\log{L_{\rm bol}/L_{\odot}} = 5.64 \pm 0.12$ and $5.53 \pm 0.08$ respectively for the primary and secondary \citep{Put23}. Comparing the X-ray luminosities, evaluated from the fluxes in the 0.5 -- 10\,keV range corrected for the interstellar absorption only, with these bolometric luminosities we find $\frac{L_{\rm X}}{L_{\rm bol}}$ of $1.5\,10^{-7}$ at apastron and $3.4\,10^{-7}$ at periastron. If we adopt instead typical bolometric magnitudes as quoted by \citet{Martins} for stars of equal spectral types, we obtain $\frac{L_{\rm X}}{L_{\rm bol}}$ of $1.2\,10^{-7}$ at apastron and $2.7\,10^{-7}$ at periastron.

In this context, it is interesting to consider the $\frac{L_{\rm X}}{L_{\rm bol}}$ ratio corresponding to the constant term in Eq.\,\ref{eq1d}. Adopting the bolometric luminosities from \citet{Put23}, we find that this constant term yields $\frac{L_{\rm X}}{L_{\rm bol}} = 1.3\,10^{-7}$. This constant X-ray emission likely reflects the intrinsic emission due to wind-embedded shocks in the two stars in the binary system. Quite remarkably the associated luminosity follows very closely the canonical relation for the intrinsic emission of O-type stars \citep{Ber95,Naze09}. From Eq.\,\ref{eq1d}, we further note that the wind-embedded shocks (i.e. the constant term in the relation) contribute between 38\% (at periastron) and 81\% (at apastron) of the total X-ray flux. 

The presence of a population of relativistic electrons in the wind interaction region of HD\,168112 further opens up the possibility that IC scattering of stellar UV photons by these relativistic electrons could result in a non-thermal X-ray or $\gamma$-ray emission \citep{Pit21}. Indeed, the $\eta$~Car colliding wind system was detected at energies above 20\,keV with the International Gamma Ray Astrophysics Laboratory (INTEGRAL) and {\it NuSTAR} \citep{Ley10,Ham18}, and up into the GeV domain with Agile and {\it Fermi} \citep{Far11,Rei15}. So far all attempts to detect such a non-thermal X-ray emission over the 0.5 -- 10\,keV band in other particle accelerating colliding wind binaries failed \citep[e.g.][]{Rau02,Mos20}. This indicates that any non-thermal X-ray emission must be significantly weaker than the thermal emission from the colliding winds binary, at least at energies below 10\,keV.

\citet{Pit21} estimated that IC energy losses by relativistic electrons are important for systems with orbital separation of less than $10^{14}$\,cm. For HD\,168112, the orbital separation varies between 248\,R$_{\odot}$ ($1.73\,10^{13}$\,cm) at periastron and 1766\,R$_{\odot}$ ($1.23\,10^{14}$\,cm) at apastron adopting the \citet{Blo24} orbital elements and taking $i=63^{\circ}$. Hence, IC scattering by relativistic electrons should be relevant for this system over most orbital phases, and certainly around periastron. Our spectral fitting tests including a power law component yielded photon indices that varied over a wide range and were usually quite different from the expected value of 1.5. In this context, it is interesting to note that \citet{Pit21} drew attention to the large variations in the spectral indices of the relativistic particles in their models of colliding wind binaries, which make the usual $N(E) \propto E^{-2}$, where $N(E)$ stands for the number of photons emitted with an energy $E$, too simplistic. In the energy range between 1 and 10\,keV, where non-thermal emission is dominated by IC scattering, their models always predict a photon index $\Gamma$ in the range 1.5 to 2. Hence, we performed another series of spectral fittings adopting a model as given by Eq.\,\ref{fitexpression_power}, but this time keeping the photon index fixed to $\Gamma = 1.5$. These fits were of significantly poorer quality than those described in Sect.\,\ref{fits168112}. Nevertheless, the normalisation of the power-law component was usually significant at the $\sim 4\,\sigma$ level and its intrinsic flux in the 0.5\,keV -- 10\,keV band would be typically of order $3\,10^{-13}$\,erg\,cm$^{-2}$\,s$^{-1}$, corresponding to $\sim 10$\% of the total intrinsic flux. However, we note that these spectral fits failed to properly reproduce the Fe\,{\sc xxv} line at 6.7\,keV, which requires a high-temperature thermal plasma (see Table\,\ref{tab:fit3THD168112}). Therefore, we conclude that the existing data do not provide any evidence for a significant non-thermal X-ray emission in HD\,168112. The above quoted flux associated with such a component is thus to be considered as a very conservative upper limit.     

\begin{figure*}
  \begin{minipage}{8.5cm}
  \begin{center}
    \resizebox{8.5cm}{!}{\includegraphics{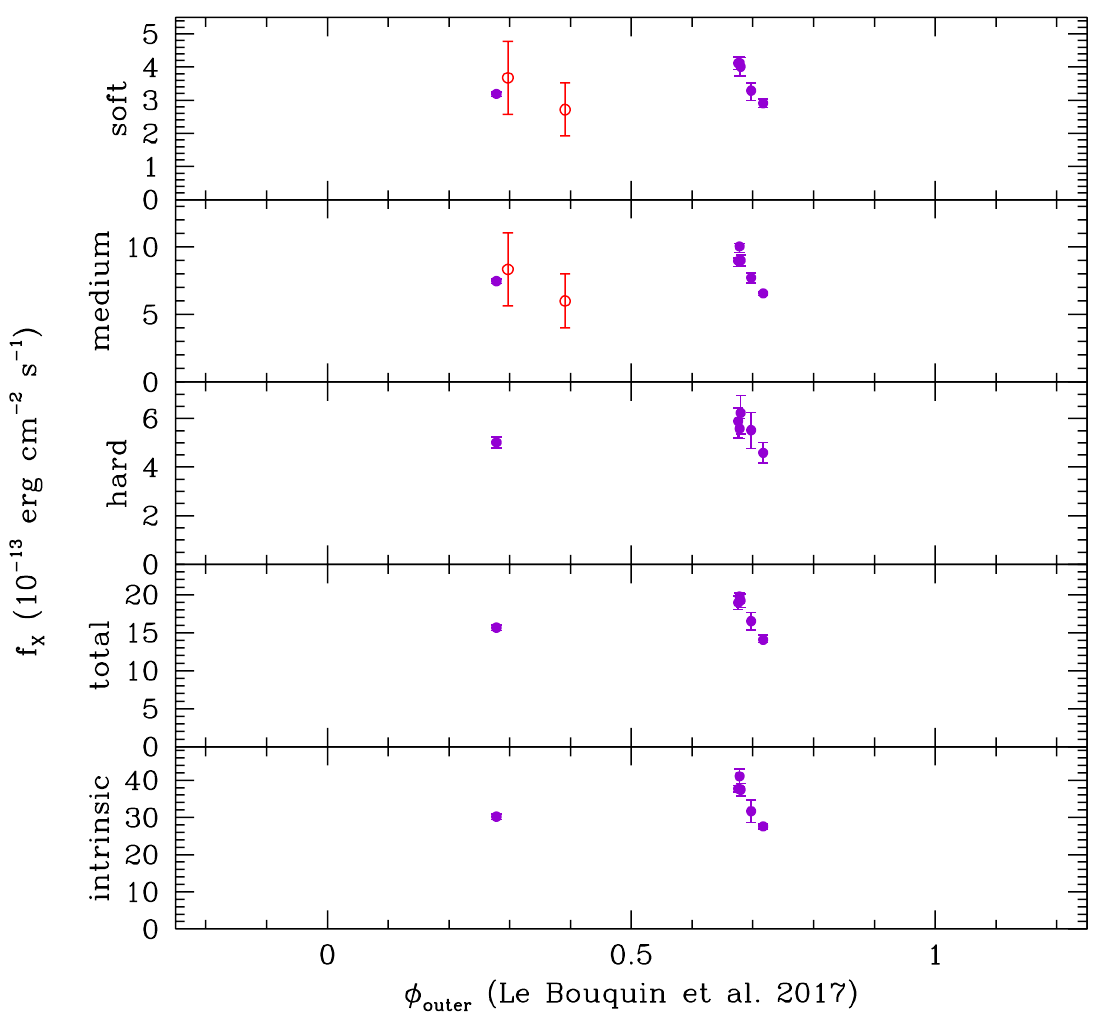}}
  \end{center}
  \end{minipage}
  \hfill
  \begin{minipage}{8.5cm}
  \begin{center}
    \resizebox{8.5cm}{!}{\includegraphics{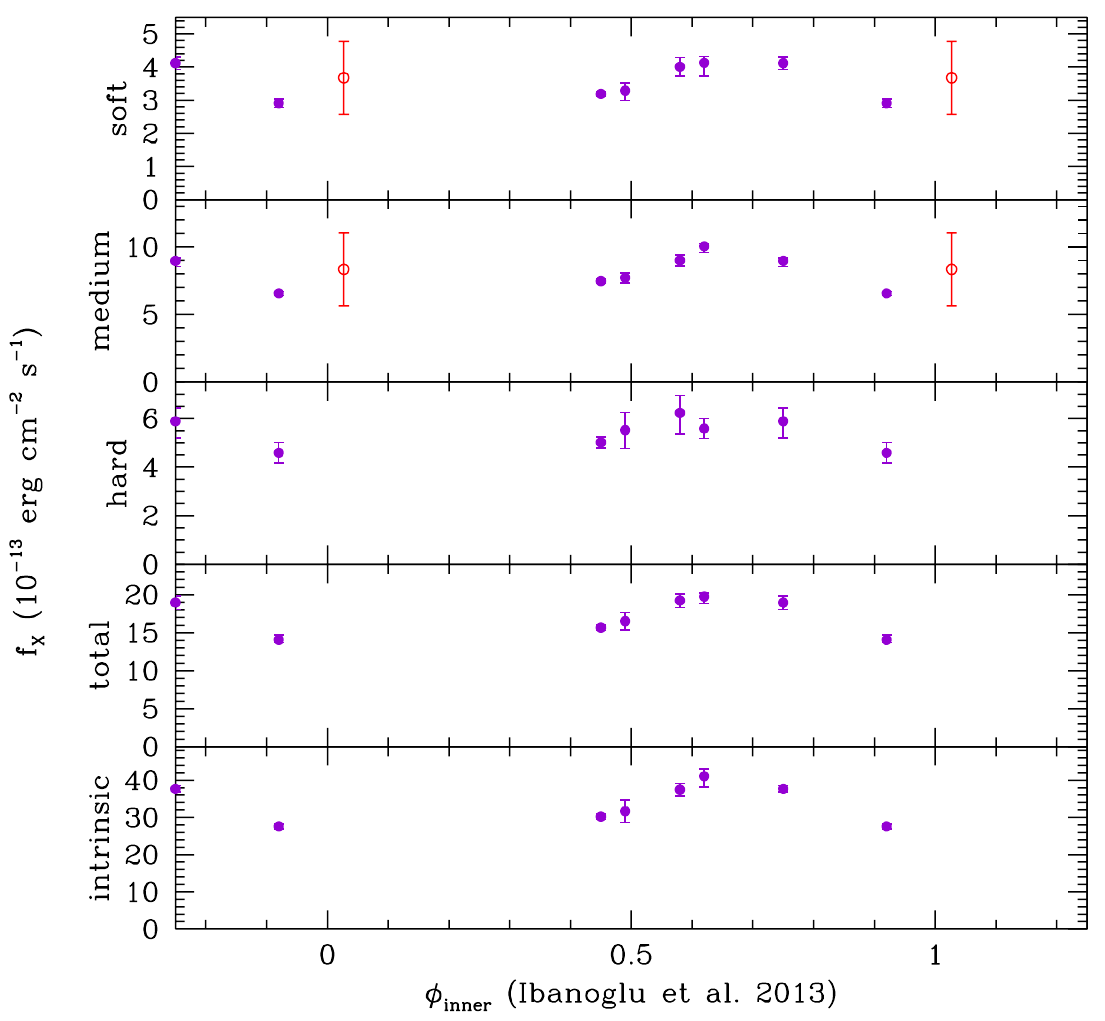}}
  \end{center}
  \end{minipage}    
  \caption{Variations in the X-ray flux of HD\,167971. Left panel: X-ray flux of HD\,167971 as a function of the orbital phase of the tertiary component \citep{LeB17}. Phase $\phi_{\rm outer}=0$ corresponds to periastron passage. The top four panels correspond to the observed X-ray flux in four different energy bands: soft (0.5\,keV -- 1.0\,keV), medium (1.0\,keV -- 2.0\,keV), hard (2.0\,keV -- 4.0\,keV), and total (0.5\,keV -- 10\,keV). Violet filled symbols indicate {\it XMM-Newton} data, and open red circles stand for ROSAT observations. The bottom panel yields the flux in the 0.5\,keV -- 10\,keV band corrected for absorption by the ISM. Right panel: Same but as a function of the orbital phase of the inner eclipsing binary \citep{Iba13}. Here $\phi_{\rm inner}=0$ corresponds to the primary minimum of the eclipsing binary. Because the ROSAT-HRI observation spanned nearly one month (i.e. more than eight orbital cycles) it is not shown here.\label{variab167971}}
\end{figure*}
\subsection{Colliding winds in HD\,167971 \label{CWB167971}}
The X-ray fluxes of HD\,167971 in the 0.5 -- 10\,keV band corrected for the interstellar absorption range between $2.75\,10^{-12}$\,erg\,cm$^{-2}$\,s$^{-1}$ and $4.17\,10^{-12}$\,erg\,cm$^{-2}$\,s$^{-1}$. The bolometric luminosities of the eclipsing binary components were evaluated as $\log{L_{\rm bol}/L_{\odot}} = 5.383$ and $5.102$, whilst the third light contribution was found to be near 40\% of the total light of the triple system \citep{Iba13}. Evaluating the sum of these bolometric luminosities, assuming a 2\,kpc distance and comparing with the above X-ray fluxes corrected for the interstellar absorption results in $\frac{L_{\rm X}}{L_{\rm bol}}$ between $3.8\,10^{-7}$ and $5.7\,10^{-7}$. Using instead the typical bolometric magnitudes as quoted by \citet{Martins} for stars of equal spectral types yields $\frac{L_{\rm X}}{L_{\rm bol}}$ between $2.0\,10^{-7}$ and $3.1\,10^{-7}$. HD\,167971 thus appears mildly overluminous in X-rays compared to the canonical $L_{\rm X}/L_{\rm bol}$ relation.

\begin{figure}
  \begin{center}
    \resizebox{8cm}{!}{\includegraphics{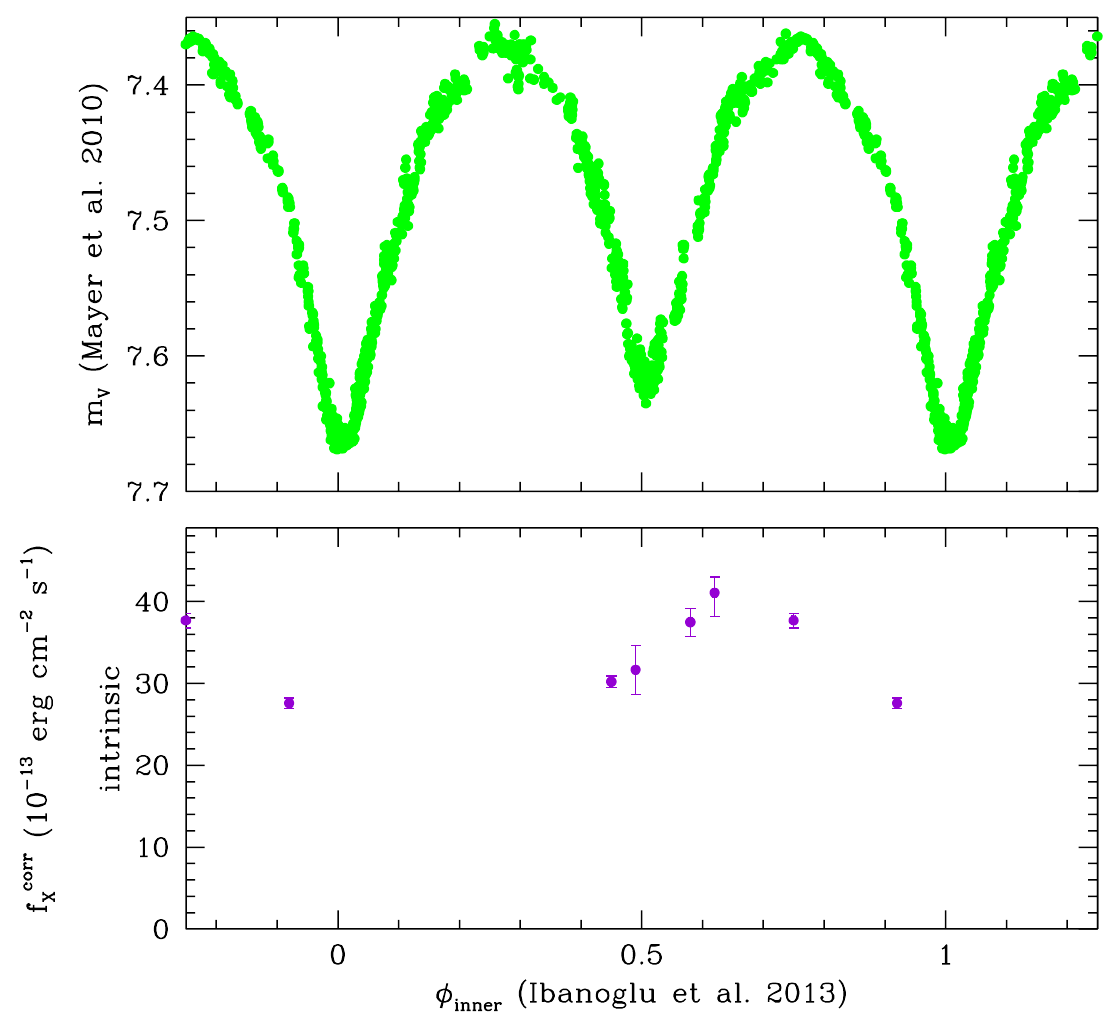}}
  \end{center}
  \caption{Comparison between optical and X-ray light curve of HD\,167971. Top panel: $V$-band photometry of HD\,167971 \citep{May10} folded with the ephemerides of the inner eclipsing binary. Bottom panel: X-ray flux in the 0.5\,keV -- 10\,keV band corrected for the interstellar absorption and folded with the ephemerides from \citet{Iba13}. We note that the uncertainties on the $\phi_{\rm inner}$ of the X-ray data are smaller than the size of the symbols. They range between $1.6\,10^{-3}$ for the oldest data and $2.8\,10^{-3}$ for the most recent observation.\label{MYSer}} 
\end{figure}

Figure\,\ref{variab167971} displays the variations in the X-ray flux as a function of the orbital phase of the tertiary component (left panel) and of the orbit of the inner eclipsing binary (right panel). The observed X-ray emission of HD\,167971 is a combination of (i) the intrinsic emission of the three components of this triple system, (ii) a possible contribution from the wind interaction zone of the short-period binary, and (iii) the collision between the combined winds of the inner binary and the wind of the third component. The third of these contributions is expected to undergo a $1/r$ modulation similar to the one observed for HD\,168112. However, Fig.\,\ref{variab167971} unveils very little coherent variations with the phase of the outer orbit. This could primarily be due to the rather scarce phase coverage. Indeed, existing {\it XMM-Newton} data only sample phases $\phi_{\rm outer}$ around 0.3 and 0.7. Because these phases are roughly symmetrical with respect to periastron ($\phi_{\rm outer} = 0.0$) and correspond to separations differing by only a few percent, the $1/r$ values are very similar and the ensuing X-ray emissions should thus display a similar level, as observed. By far, the largest variations in the X-ray emission from the outer colliding wind interaction are expected around periastron, but such data are currently lacking. The data points collected around $\phi_{\rm outer} = 0.7$ display significant variations, but this variability occurs with a timescale that is too short to be due to a $1/r$ modulation. The right panel of Fig.\,\ref{variab167971} shows instead that these variations probably arise on the timescale of the inner orbit. Indeed, the existing data suggest a modulation that is reminiscent of eclipses in the inner binary (see Fig.\,\ref{MYSer}). This is especially true for the soft and medium energy bands, whilst the variations in the hard band are more erratic. In the medium energy band, the variations have peak-to-peak amplitudes of 40\%.

The part of HD\,167971's X-ray emission that undergoes a short-term modulation consistent with the orbital period of the inner eclipsing binary must therefore arise inside this short period binary. There are three possibilities to explain the origin of this emission:
\begin{itemize}
\item[-] It could arise from the intrinsic emission due to wind-embedded shocks \citep{Feld}. The modulation of the X-ray flux would then result from eclipses by the stars and their winds, as was suggested, for instance, for the case of 29\,CMa \citep{Ber95}. Observationally, the intrinsic emission of O-type stars typically amounts to $10^{-7}\,L_{\rm bol}$ and is mostly soft with $kT \sim 0.6$\,keV \citep[e.g.][]{Ber96,Naze09}. Based on the spectral types proposed by \citet{Mai19}, and the calibration of \citet{Martins}, we estimate $\log{(L_{\rm bol}/L_{\odot})} \simeq 5.91$ for the primary and $\log{(L_{\rm bol}/L_{\odot})} \simeq 5.69$ for the secondary component. Assuming a distance of 2\,kpc, the X-ray flux due to wind-embedded shocks is estimated as $\sim 6.5\,10^{-13}$\,erg\,s$^{-1}$\,cm$^{-2}$ for the primary and $\sim 3.9\,10^{-13}$\,erg\,s$^{-1}$\,cm$^{-2}$ for the secondary. Each star contributes a non-negligible fraction ($\sim 15$\% and $\sim 9$\% respectively for the primary and secondary star) of the maximum total ISM-corrected X-ray emission, but these contributions fall short by about a factor of 2 to 4 compared to what is needed to explain the observed $\sim 40$\% peak to peak variations by occultation and eclipse effects. 
\item[-] It could arise from a wind interaction between the components of the inner binary. Indeed, short-period massive binaries may display a phase shift in their X-ray light curve with respect to the optical light curve. This is due to the Coriolis deflection of the wind interaction region (e.g. V444\,Cyg, $P_{\rm orb} = 4.21$\,d, \citealt{Lom15}; WR\,21 and WR\,31, $P_{\rm orb} = 8.25$\,d and 4.83\,d, \citealt{Naz23}). \citet{FdP15} argued that because of the radiative nature of the wind interaction in the inner binary, the total X-ray emission of HD\,167971 should be dominated by this contribution. However, from the observational viewpoint, wind interactions in such short-period O-star binaries are often X-ray faint (e.g. 29\,CMa; \citealt{Ber95}, and HD\,149404, \citealt{Rau24}). From a theoretical viewpoint, radiative wind interaction zones are expected to be subject to thin shell instabilities that considerably lower their X-ray emission \citep{SBP,Kee}. The problem would be even worse if the inner eclipsing binary is in an overcontact configuration as suggested by \citet{May10} and \citet{Iba13} based on their photometric light curve analyses. In such cases, the strong head-on collision of the winds along the line of centres is lacking and the wind interaction region has an annular shape, which would make it less prone to eclipsing effects \citep{Mon13}. Whilst we cannot rule out the colliding winds as the origin of the emission, a full hydrodynamic modelling of the wind interaction is probably needed to assess the plausibility of this scenario.
\item[-] It could form in a magnetically confined wind of one of the components of the inner binary. In magnetic massive stars, the mostly dipolar magnetic field channels the wind material towards the magnetic equator where the flows from the two hemispheres collide, leading thereby to an enhanced X-ray emission \citep[for a review see][]{udD16}. The eclipse of such a magnetically confined wind by the companion star could indeed lead to a modulation of the observed X-ray emission. \citet{Hubrig} reported the detection of a strong ($1324 \pm 582$\,G) longitudinal magnetic field in one of the components of the eclipsing binary. However, given the large error bar, this 2.3\,$\sigma$ detection requires confirmation. We note that \citet{Neiner} had previously analysed the same observation, reporting a non-detection. Again, more evidence needs to be gathered to check the validity of this scenario.
\end{itemize}

Whilst the above discussion highlights the fact that the X-ray emission of HD\,167971 arises from several regions, one obviously expects a contribution coming from the interaction between the winds of the inner binary and the wind of the tertiary component. In fact, the non-thermal radio emission of this system is clearly ruled by the outer orbit \citep{Blo07}, pointing towards an energetic wind-wind collision. \citet{Blo07} estimated that the tertiary star was at its closest position to the inner binary in 1988 in good agreement with the ephemerides of \citet{LeB17}\footnote{This is also confirmed by the EVN radio data of \citet{FdP24} that were collected near apastron of the tertiary component at a time when the synchrotron emission of this system was at a low level. The radio emission of HD\,167971 remained unresolved in this observation.}. \citet{San19} discussed two very long baseline interferometry (VLBI)  observations in 2006 and 2016. These data unveiled a change in orientation of the elongated emission region, consistent with the orbital motion of the tertiary component. These results demonstrate that the non-thermal radio emission from HD\,167971 arises from the wind-wind collision between the inner eclipsing binary and the tertiary component. The location of the emitting region between the positions of the inner binary and the tertiary further indicates that the combined wind of the inner binary is stronger than the wind of the tertiary star. The next periastron passage of the tertiary component should take place in 2030. From the $0.44 \pm 0.02$ eccentricity derived by \citet{LeB17}, one can then estimate that the contribution of the outer colliding wind interaction should increase by more than a factor of 2 between the existing data around $\phi_{\rm outer} \simeq 0.3$ or $\simeq 0.7$ and periastron if there is no impact of shock modification.

The VLBI data of \citet{San19} yielded a radio spectral index of $\alpha = -1.1$, significantly steeper than expected for synchrotron emission in an optically thin environment ($\alpha = -0.5$). \citet{San19} interpreted this as a consequence of efficient IC cooling leading to a softening of the electron energy spectrum or as a consequence of modified shocks. Whilst the current sampling of the outer orbit is insufficient to search for evidence for or against shock modification, we tried to look for evidence of non-thermal X-ray emission arising from the IC scattering by the relativistic electrons. As for HD\,168112, we repeated the fits of a model including beside a 2-T thermal plasma a power-law component with $\Gamma$ fixed to 1.5. These fits were again of poorer quality than those described in Sect.\,\ref{fits167971}. The normalisation of the power-law was only significant at the $\sim 3.5\,\sigma$ level. The intrinsic flux in the 0.5\,keV -- 10\,keV band of the power-law component would be typically of order $4\,10^{-13}$\,erg\,cm$^{-2}$\,s$^{-1}$, corresponding to $\sim 3$\% of the total ISM-corrected X-ray flux. We stress that these models failed to reproduce the Fe\,{\sc xxv} line at 6.7\,keV clearly seen in the EPIC-pn spectrum from observation 3. Hence, as in the case of HD\,168112, we conclude that the flux quoted for the non-thermal X-ray emission must be seen as a strict upper limit and that currently, there is no detection of a genuine non-thermal high-energy emission in HD\,167971.     

\section{Conclusions}
In this study we used a new set of {\it XMM-Newton} observations to gain further insight into the X-ray properties of two colliding wind systems that are well-known non-thermal radio emitters. Taking advantage of the orbital elements that were recently established for HD\,168112, we find that the X-ray spectra of this highly eccentric 1.4\,yr binary unveil a clear phase-dependent X-ray over-luminosity. The X-ray fluxes display variability consistent with the expected $1/r$ modulation for an adiabatic wind interaction. We show that the shocks remain adiabatic at periastron and do not collapse, unlike what has been observed in other eccentric massive binaries. Despite the presence of a population of relativistic electrons, as revealed by the synchrotron radio emission, we find no evidence of a significant shock modification due to the action of these relativistic electrons. We also failed to detect any clear indication of a non-thermal X-ray emission that could arise from IC scattering by relativistic electrons. Our analysis indicates that wind-embedded shocks in the individual winds of the binary contribute between 38\% and 81\% of the X-ray flux at periastron and apastron, respectively.   

The existing X-ray data of the triple system HD\,167971 were not sufficient to look for a $1/r$ modulation locked to the orbit of the tertiary component. This is mostly because observations of the outer orbit near periastron are still missing. From the current best knowledge of the outer orbit, the next opportunity to fill this gap is expected around 2030. Meanwhile, the existing data hint at a modulation of the X-ray emission on the orbital period of the inner eclipsing binary. This suggests that a significant fraction of the observed X-ray emission must arise inside this inner binary, although its exact origin remains uncertain. Whilst the emission from wind-embedded shocks is probably not sufficient to explain the observed modulation, colliding winds possibly coupled with a magnetically confined wind of one of the components of the inner binary could provide an explanation. However, a better sampling of the inner orbital cycle would be needed to determine the constraints needed to guide future models of this system.  

\begin{acknowledgements}
The Li\`ege team acknowledges support from the Fonds National de la Recherche Scientifique (Belgium) and the Belgian Federal Science Policy Office (BELSPO) in the framework of the PRODEX Programme (HERMeS contract). ADS and CDS were used for this research. 
\end{acknowledgements}

\end{document}